\documentclass[twocolumn,floats,showpacs,amsmath,amssymb,prl]{revtex4}
\usepackage{graphicx}
\begin{document}

\title{Diffusion of muonium and hydrogen in diamond}
\author{Carlos P. Herrero}
\author{Rafael Ram\'{\i}rez}
\affiliation{Instituto de Ciencia de Materiales de Madrid,
         Consejo Superior de Investigaciones Cient\'{\i}ficas (CSIC),
         Cantoblanco, 28049 Madrid, Spain }
\date{\today}

\begin{abstract}
 Jump rates of muonium and hydrogen in diamond are calculated by quantum 
transition-state theory, based on the path-integral centroid formalism.
This technique allows us to study the influence of vibrational mode 
quantization on the effective free-energy barriers $\Delta F$ for impurity 
diffusion, which are renormalized respect to the zero-temperature classical
calculation. For the transition from a tetrahedral (T) site to a bond-center
(BC) position, $\Delta F$ is larger for hydrogen than for muonium, and the 
opposite happens for the transition BC $\to$ T.
The calculated effective barriers decrease for rising temperature, except for 
the muonium transition from T to BC sites. 
Calculated jump rates are in good agreement to available muon spin rotation
data.
\end{abstract}


\maketitle
Hydrogen has been studied for many years as an impurity in solids.
Due to its low mass, this impurity has posed some challenging problems 
to both experimentalists and theorists \cite{es95}.
An important property of hydrogen in insulating and semiconducting
materials is its ability to form complexes and passivate defects
\cite{pe92}, as was found in the case of diamond \cite{ze99}.
A large amount of information has been obtained by studying muonium
(formed by a muon $\mu^+$ and an electron), which behaves as a light
isotope of hydrogen, with $m_{\mu} \approx m_p / 9$. In particular,
muon spin rotation ($\mu$SR) experiments have provided us with invaluable
information on the behavior of muonium (Mu) in semiconductors \cite{pa88}.

An interesting topic is the diffusion of Mu and H in crystalline
materials. This problem turns out to be difficult due to the combination
of quantum effects with lattice relaxation around the impurity 
(polaron effect). In fact,
zero-point motion influences the excitation energy from the ground state 
to the top of the diffusion barrier. Also, tunneling of the impurities 
can be enhanced by phonons, and lattice distorsions 
may depend on the isotopic mass of the impurity.

Two types of isolated Mu centers have been observed in elemental 
semiconductors, and are characterized by their isotropic or anisotropic 
hyperfine interaction. The former is accepted to consist of Mu at
an interstitial tetrahedral (T) site, while the latter corresponds to Mu
at a bond-center (BC) site, midway between two nearest host atoms.
Muon implantation experiments in diamond, as well as various theoretical
approaches, have shown that this impurity is metastable at the T site,
and has its lowest energy at or around the BC site,
as a result of a large lattice relaxation \cite{es87,es95,go03}.
An important fraction of the implanted muons form
Mu$_{\text T}$, and the transition to 
Mu$_{\text {BC}}$ has been observed by $\mu$SR \cite{pa88}.
This transition T $\to$ BC, and the opposite BC $\to$ T are also expected
to be important for the diffusion of hydrogen in diamond, as derived from
the energy barriers calculated with several theoretical methods \cite{go02}.

In diamond there is the additional problem that H and Mu may exist in 
different charge states, and moreover a change of state may occur in 
combination with hopping.
Density functional (DF) theory calculations predict H$^+$ to be stable 
off-axis in a buckled bond-centered configuration,
in contrat to the BC site for neutral H \cite{go02,go03}.
Also, calculated diffusion barriers were found to change appreciably
with the impurity charge state \cite{go03,go02}.
Here we will concentrate on high-resistivity diamond, where the
non-paramagnetic fraction does not play an important role \cite{es95},
and thus we will study neutral H and Mu.

In earlier works, hydrogen diffusion in diamond has been studied 
theoretically in the classical (high-temperature) limit \cite{go03,ka00}.
However, quantum effects are important for hydrogen-related defects 
in this material at relatively low temperatures \cite{ke04}. 
Thus, we study in this Letter the diffusion of H and Mu in diamond 
by explicitly considering the impurities as quantum particles.
The main question to be answered is the dependence of hopping rates on
both temperature and impurity mass.  In this respect,
there is a vast literature about theoretical models for quantum diffusion 
of light particles in solids, and metals in particular 
\cite{fl70}.
Due to the complexity of this problem, such computations have been
typically based on model potentials for the impurity-lattice interactions.

Here we calculate the jump rate of hydrogen and muonium by 
quantum transition-state theory \cite{gi87b}, 
using path-integral molecular dynamics (PIMD) simulations.  
We employ a realistic interatomic potential, derived from
DF theory calculations.
This method allows us to calculate jump rates for this nonlinear 
many-body problem, including lattice relaxation, 
zero-point motion, and phonon-assisted incoherent tunneling. 

 In the path-integral formalism of statistical mechanics, a
quantum particle can be represented as a cyclic chain of $L$ beads
coupled by harmonic springs ($L$, Trotter number).
This formalism has been employed earlier to study equilibrium properties of
H and Mu in silicon \cite{ra94} and diamond \cite{he06},
by using Monte Carlo simulations.
In this context, there exists a quantum extension of classical 
transition-state theory for calculating rate constants of infrequent 
events \cite{gi87b}.
 It relates the jump rate $k$ to the probability 
density of the center-of-gravity
(centroid) of the quantum paths of the jumping impurity,
and particularly to the ratio $P_c$ between the equilibrium probability 
of finding the centroid at a saddle-point (say ${\bf r}^*$) and at a stable 
site (say ${\bf r}_0$) \cite{gi87b}.  Namely:
$ k = \overline{v}  P_c / 2 l$, where $l$ is the distance between 
${\bf r}_0$ and ${\bf r}^*$ and $\overline{v}$ is a weakly
temperature-dependent factor: 
$\overline{v} = 2 \sqrt{\pi} \, \phi \, \Delta x / (\beta \hbar)$.
Here, $\Delta x$ is the width of the probability distribution for the 
jumping impurity with its 
centroid $\overline{{\bf x}}$ fixed at ${\bf r}^*$, 
$\beta = 1/k_B T$, and $\phi$ is a number of order one at low $T$. 
 $P_c$ can be written as $\text{exp}(-\beta \Delta F)$, $\Delta F$ being 
an effective free-energy barrier, given by the reversible work done on the 
system when the impurity centroid $\overline{{\bf x}}$ moves along a path 
from ${\bf r}_0$ to ${\bf r}^*$:
\begin{equation}
 \Delta F = - \int_{{\bf r}_0}^{{\bf r}^*} {\bf f}(\overline{{\bf x}})
     d\overline{{\bf x}}  \, ,
\label{pc}
\end{equation}
where ${\bf f}(\overline{\bf x})$ is the mean force acting 
on the impurity with its centroid fixed on $\overline{{\bf x}}$
at temperature $T$:
$ {\bf f}(\overline{{\bf x}}) = - \langle \nabla_{\bf x} V({\bf R})
                  \rangle_{\overline{\bf x}}$.
Here $V({\bf R})$ is the potential energy, ${\bf R}$ being in our case a 
$3(N+1)$-dimensional vector ($N$ host atoms plus one impurity).
The reliability of this method to calculate free-energy barriers and 
jump rates was discussed in \cite{gi87b,ma95}.

We use the Born-Oppenheimer approximation to define a potential
energy surface $V({\bf R})$ for the nuclear coordinates.  Since
true {\em ab-initio} potentials
require computer resources that would enormously restrict the size of
the simulation cell, we have found a compromise by
using an efficient tight-binding (TB) Hamiltonian, 
based on DF calculations \cite{po95}.
With this interaction potential we found the BC site to be
the absolute energy minimum for H \cite{he06}, and the energy surface 
is similar to that derived from earlier DF and TB calculations \cite{go03}

To sample the configuration space we employed the PIMD method
in the $NVT$ ensemble \cite{ma96,he06}.
Simulations were carried out on a $2 \times 2 \times 2$ supercell 
of the diamond face-centered cubic cell with periodic boundary conditions,
including 64 C atoms and one impurity.
To assure the right convergence of the path integrals, we took a Trotter 
number $L \propto 1/T$, with $L$ = 20 for H and 60 for Mu at 300 K.
For given impurity mass and temperature, 
the mean force ${\bf f}(\overline{{\bf x}})$ was calculated 
at 14 points along the integration line between ${\bf r}_0$ and 
${\bf r}^*$ in Eq.~(\ref{pc}). 
 For a centroid position $\overline{{\bf x}}$, the nearest and next-nearest 
neighbors of the impurity (up to a total of 13 host atoms) are treated 
quantum-mechanically to obtain ${\bf f}(\overline{{\bf x}})$, as in
\cite{he97}. 
For each point in the integration paths, we generated 5000 configurations 
for system equilibration, and 3 $\times$ $10^4$ configurations to
calculate ensemble average properties. More technical details 
are given elsewhere \cite{ma96,he06}.

\begin{figure}
\vspace*{-1.7cm}
\includegraphics[width=8.5cm]{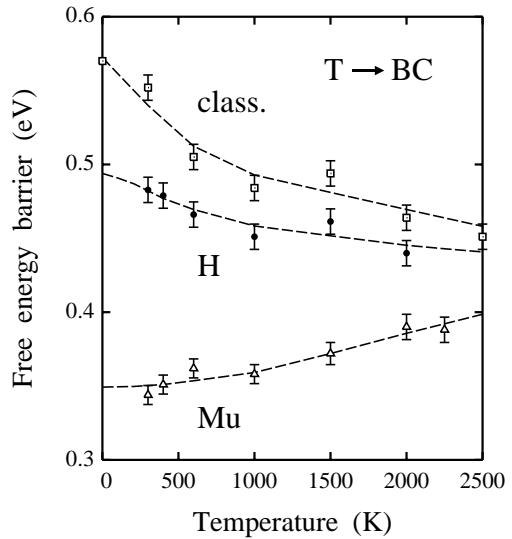}
\vspace{-1.8cm}
\caption{
Effective free-energy barrier for impurity jumps from a T site
to a neighboring BC site as a function of temperature. 
Open triangles, muonium;
filled circles, hydrogen; open squares, classical limit.
Dashed lines are guides to the eye.
}
\label{f1}
\end{figure}

In Fig.~\ref{f1} we present the free-energy barrier $\Delta F$ 
for impurity diffusion
from a T to a BC site. Data derived from line integration of the mean force 
are shown as a function of temperature for hydrogen (solid circles) and 
muonium (triangles). In this plot, one notices first that $\Delta F$
is higher for H than for Mu. 
At low temperature, the dependence of $\Delta F$ upon impurity mass
is related to the change in internal energy $E$ of the
defect complex along the diffusion path. 
To compare Mu and H, we note that $E_{\text{Mu}}$ is always larger 
than $E_{\text H}$, but the difference  
$E_{\text{Mu}} - E_{\text H}$ changes from 0.96 eV at a site T to 0.83 eV 
at the transition state, thus giving 
$(\Delta F)_{\text H} - (\Delta F)_{\text{Mu}} = 0.13 \pm 0.01$ eV.
Second, in the case of H, one observes a slight decrease in $\Delta F$ 
for increasing $T$, contrary to the rise in effective free-energy 
barrier for muonium. At high $T$, one converges to the other, as expected
for the classical limit.
These migration barriers are on the order of that obtained
earlier from DF-TB calculations in Ref.~\onlinecite{ka00} ($0.4 \pm 0.1$ eV).

\begin{figure}
\vspace*{-1.7cm}
\includegraphics[width=8.5cm]{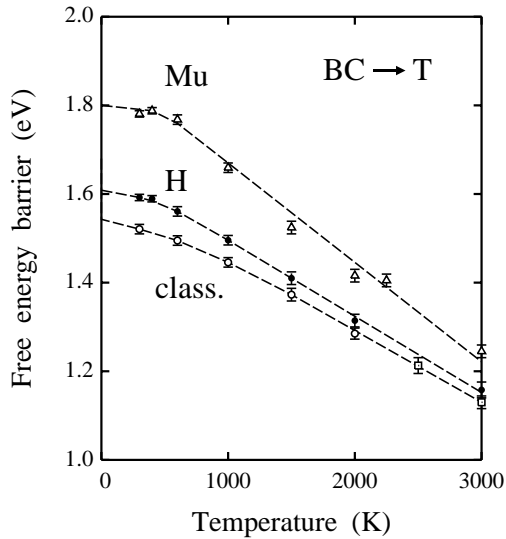}
\vspace{-1.8cm}
\caption{
Free-energy barrier for impurity jumps from a BC site to a neighboring
tetrahedral T site as a function of temperature.
Triangles, muonium; filled circles, hydrogen; open circles,
classical limit. Dashed lines are guides to the eye.
}
\label{f2}
\end{figure}

For the opposite jump (impurity from BC to T), the effective barrier is higher 
for Mu than for H. This is shown in Fig.~\ref{f2}, where symbols correspond
to $\Delta F$ derived from our PIMD simulations at several temperatures. 
(Note the different vertical scales in Figs.~\ref{f1} and ~\ref{f2}.)
In this case, $E_{\text{Mu}} - E_{\text H}$ changes at low temperature 
from 0.64 eV at a BC site to 0.83 eV at the transition state, giving:
$(\Delta F)_{\text H} - (\Delta F)_{\text{Mu}} = -0.19 \pm 0.01$ eV.
The low-temperature energy barriers obtained here for H and Mu are
comparable to those found from DF theory (1.6 eV) \cite{go02}
and earlier DF-TB calculations ($2.0 \pm 0.1$ eV) \cite{ka00}.
Our simulations yield, however, an important decrease in $\Delta F$  
as temperature is raised.

This impurity transition between BC and T sites is asymmetric in a double 
sense. First, it happens between a local minimum of the energy surface
and the absolute energy minimum.  Second, the lattice
relaxations involved in both impurity positions are very different.
At BC, the nearest host atoms relax strongly ($\sim 0.4$ \AA), whereas 
at a T site the relaxation of the C atoms is much weaker ($0.08$ \AA).
This second asymmetry is relevant for the temperature dependence of the
free-energy barriers shown in Figs.~\ref{f1} and \ref{f2}.
For diffusion across static barriers, it is known that $\Delta F$ 
increases with temperature \cite{ra97b}. 
In our case, the barrier $\Delta F$ depends on lattice ralaxation and
vibrational modes, and the larger the relaxation, the more inportant is
the change in $\Delta F$ with temperature.
For the transition BC $\to$ T, $\Delta F$ is controlled by the energy surface
around BC, in which the lattice relaxation changes appreciably from
the BC site to the transition point, and thus $\Delta F$ decreases
fast for rising $T$.
On the contrary, $\Delta F$ for the jump T $\to$ BC is controlled
by the path between T and the transition point, which does not involve
important host-atom relaxations, and $\Delta F$ changes slowly with $T$.
In particular for increasing $T$, $\Delta F$ decreases for H but rises
for Mu, as a consequence of the smaller lattice relaxation for Mu
(closer to a static barrier). 

\begin{figure}
\vspace*{-1.7cm}
\includegraphics[width=8.5cm]{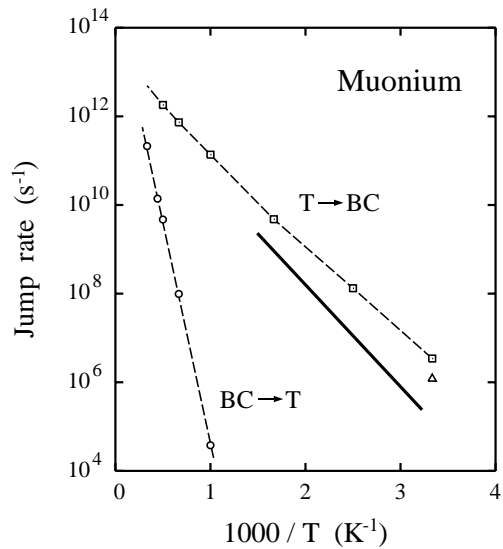}
\vspace{-1.8cm}
\caption{
 Rate for impurity jumps from a T site to a BC site.
Open squares represent results derived from PIMD calculations.
The solid line corresponds to data derived from $\mu$SR
measurements \cite{od88}, and an open triangle indicates
the rate measured in \cite{ma00}.
For comparison, we also present PIMD results for the jump rate from
BC to T (circles).
Error bars of the simulation data are on the order of the symbol size.
Dashed lines are guides to the eye.
}
\label{f3}
\end{figure}

In Fig.~\ref{f3} we show the muonium jump rate from a T site to a 
neighboring BC site, as derived from the probability $P_c$ given by the
calculated barriers $\Delta F$ (open squares).
One observes a certain departure from linearity
in the Arrhenius plot, due basically to the change in
effective barrier as a function of temperature (see Fig.~\ref{f1}). 
The solid line in Fig.~\ref{f3} displays the transition rate derived by 
Odermatt {\em et al.} \cite{od88} from $\mu$SR experiments.
A triangle shows the spin relaxation rate of Mu$_{\text T}$ in
insulating diamond containing vacancies,
and measured at room temperature \cite{ma00}.
In that work, it was found a rather constant relaxation rate at $T <$ 100 K,
which could indicate the appearance of coherent tunneling between T sites.
For comparison, we also give in Fig.~\ref{f3} the Mu jump rate from 
BC to T sites, as derived from the free-energy barriers shown in Fig.~\ref{f2}.
At room temperature, and even at $T \sim$ 1000 K, this jump rate is
several orders of magnitude smaller than that for the
transition T $\to$ BC.
From the jump rate derived from our simulations at 300 K 
($k_{\text{Mu}} = 3.4 \times 10^6 \ {\text s}^{-1}$), one expects that a muon 
implanted in diamond at a site T will likely diffuse to a BC site 
before decaying (with mean lifetime $\tau_{\mu} = 2.2 \ \mu$s).

We now turn to the diffusion of H in the diamond bulk. 
There are a number of possible migration trajectories, as suggested
by earlier theoretical works \cite{go03}.
For neutral hydrogen, according to the energy surface derived in
our calculations, one can think of two main diffusion paths for hydrogen.
The first one (denoted as path I) consists of a jump from BC to a 
neighboring T site, followed
by a transition to another BC site. This is the process envisioned in 
Ref.~\onlinecite{ka00} from (classical) locally-activated Monte Carlo 
simulations. For both atomic jumps, we obtain the effective free-energy
barriers shown in Figs.~\ref{f1} and \ref{f2}. 
An alternative path (called II) moves H from a BC site 
to a nearest BC site, having a transition state at a point with C$_{2v}$
symmetry close to the so-called $C$ site \cite{go02}. 
This path is similar to that employed earlier in a calculation
of the jump rate of H in silicon \cite{he97},
but in diamond the energy barrier is much higher \cite{wa89}. 
In fact, for classical point atoms at 
$T = 0$ we find a barrier of $2.1 \pm 0.1$ eV, close to 1.8 eV
obtained from DF calculations \cite{go02} and 1.9 eV derived
from semi-empirical cluster calculations \cite{me92}.
Concerning these energy barriers, it is worth noting that 
seemingly simple atomic jumps can actually involve coupled barriers, 
as clearly indicated in Ref.~\onlinecite{ra96}.
Thus, to obtain the barrier in path II we considered a coupled motion 
of H and the C atom lying between both BC sites.

Going to the results of our finite-temperature simulations of H 
in diamond, we find for path II at 1000 K a jump rate from BC to BC sites 
of 160 s$^{-1}$, 
much lower than that obtained in path I for the transition from BC to T 
($k = 2.5 \times 10^5$ s$^{-1}$).
Taking into account the rate for the opposite process T $\to$ BC 
($k = 4.6 \times 10^{10}$ s$^{-1}$), and the relative residence time of
hydrogen at BC and T sites, we obtain at 1000 K a diffusion coefficient 
along path I of $D_{\text H} = 3.1 \times 10^{-12}$ cm$^2$ s$^{-1}$.
This indicates that H diffusion from BC to BC sites will mainly
happen via short visits of tetrahedral T sites, or their associated 
attraction basins in configuration space.

Experimental investigations on hydrogen diffusion in diamond have been 
so far scarce. Popovici {\em et al.}~\cite{po95b} studied the diffusion
of several species in diamond at 1130 K, and found that the
diffusion coefficients for N, O, and H are very similar.
They concluded that the diffusion process of these impurities is probably
affected by crystal defects that would trap the diffusing species. 
Hence the value for hydrogen found by these
authors [$D_{\text H} = (2.4 \pm 0.3) \times 10^{-13}$ cm$^2$ s$^{-1}$] 
has to be considered as a lower limit for diffusion in a perfect diamond 
crystal.

In summary, path-integral molecular dynamics simulations provide us with 
a good tool to study quantum effects on the jump-rate of muonium and hydrogen 
in diamond. Renormalization of the classical diffusion barriers due to
these effects is appreciable. In particular, we have found that the
effective free-energy barrier for muonium can be smaller or larger than
that for hydrogen, depending on the diffusion process under consideration.
For muonium, we find jump rates from T to BC sites on the order of 
$\mu$SR data. For hydrogen, the most probable diffusion path involves BC
and T sites.

These calculations were performed at the Barcelona
Supercomputing Center (BSC-CNS). This work was supported by M.E.C.
(Spain) through Grant FIS2006-12117-C04-03.
E.R. Hern\'andez is thanked for helpful discussions.

\bibliographystyle{apsrev}

\begin{thebibliography}{24}
\expandafter\ifx\csname natexlab\endcsname\relax\def\natexlab#1{#1}\fi
\expandafter\ifx\csname bibnamefont\endcsname\relax
  \def\bibnamefont#1{#1}\fi
\expandafter\ifx\csname bibfnamefont\endcsname\relax
  \def\bibfnamefont#1{#1}\fi
\expandafter\ifx\csname citenamefont\endcsname\relax
  \def\citenamefont#1{#1}\fi
\expandafter\ifx\csname url\endcsname\relax
  \def\url#1{\texttt{#1}}\fi
\expandafter\ifx\csname urlprefix\endcsname\relax\def\urlprefix{URL }\fi
\providecommand{\bibinfo}[2]{#2}
\providecommand{\eprint}[2][]{\url{#2}}

\bibitem[{\citenamefont{Estreicher}(1995)}]{es95}
\bibinfo{author}{\bibfnamefont{S.~K.} \bibnamefont{Estreicher}},
  \bibinfo{journal}{Mater. Sci. Eng.} \textbf{\bibinfo{volume}{R14}},
  \bibinfo{pages}{319} (\bibinfo{year}{1995}).

\bibitem[{\citenamefont{Pearton et~al.}(1992)\citenamefont{Pearton, Corbett,
  and Stavola}}]{pe92}
\bibinfo{author}{\bibfnamefont{S.~J.} \bibnamefont{Pearton}},
  \bibinfo{author}{\bibfnamefont{J.~W.} \bibnamefont{Corbett}},
  \bibnamefont{and} \bibinfo{author}{\bibfnamefont{M.}~\bibnamefont{Stavola}},
  \emph{\bibinfo{title}{Hydrogen in Crystalline Semiconductors}}
  (\bibinfo{publisher}{Springer}, \bibinfo{address}{Berlin},
  \bibinfo{year}{1992}).

\bibitem[{\citenamefont{Zeisel et~al.}(1999)\citenamefont{Zeisel, Nebel, and
  Stutzmann}}]{ze99}
\bibinfo{author}{\bibfnamefont{R.}~\bibnamefont{Zeisel}},
  \bibinfo{author}{\bibfnamefont{C.~E.} \bibnamefont{Nebel}}, \bibnamefont{and}
  \bibinfo{author}{\bibfnamefont{M.}~\bibnamefont{Stutzmann}},
  \bibinfo{journal}{Appl. Phys. Lett.} \textbf{\bibinfo{volume}{74}},
  \bibinfo{pages}{1875} (\bibinfo{year}{1999}).

\bibitem[{\citenamefont{Patterson}(1988)}]{pa88}
\bibinfo{author}{\bibfnamefont{B.~D.} \bibnamefont{Patterson}},
  \bibinfo{journal}{Rev. Mod. Phys.} \textbf{\bibinfo{volume}{60}},
  \bibinfo{pages}{69} (\bibinfo{year}{1988}).

\bibitem[{\citenamefont{Estle et~al.}(1987)\citenamefont{Estle, Estreicher, and
  Marynick}}]{es87}
\bibinfo{author}{\bibfnamefont{T.~L.} \bibnamefont{Estle}},
  \bibinfo{author}{\bibfnamefont{S.}~\bibnamefont{Estreicher}},
  \bibnamefont{and} \bibinfo{author}{\bibfnamefont{D.~S.}
  \bibnamefont{Marynick}}, \bibinfo{journal}{Phys. Rev. Lett.}
  \textbf{\bibinfo{volume}{58}}, \bibinfo{pages}{1547} (\bibinfo{year}{1987}).

\bibitem[{\citenamefont{Goss}(2003)}]{go03}
\bibinfo{author}{\bibfnamefont{J.~P.} \bibnamefont{Goss}}, \bibinfo{journal}{J.
  Phys.: Condens. Matter} \textbf{\bibinfo{volume}{15}}, \bibinfo{pages}{R551}
  (\bibinfo{year}{2003}).

\bibitem[{\citenamefont{Goss et~al.}(2002)\citenamefont{Goss, Jones, Heggie,
  Ewels, Briddon, and \"Oberg}}]{go02}
\bibinfo{author}{\bibfnamefont{J.~P.} \bibnamefont{Goss}},
  \bibinfo{author}{\bibfnamefont{R.}~\bibnamefont{Jones}},
  \bibinfo{author}{\bibfnamefont{M.~I.} \bibnamefont{Heggie}},
  \bibinfo{author}{\bibfnamefont{C.~P.} \bibnamefont{Ewels}},
  \bibinfo{author}{\bibfnamefont{P.~R.} \bibnamefont{Briddon}},
  \bibnamefont{and} \bibinfo{author}{\bibfnamefont{S.}~\bibnamefont{\"Oberg}},
  \bibinfo{journal}{Phys. Rev. B} \textbf{\bibinfo{volume}{65}},
  \bibinfo{pages}{115207} (\bibinfo{year}{2002}).

\bibitem[{\citenamefont{Kaukonen et~al.}(2000)\citenamefont{Kaukonen, Perajoki,
  Nieminen, Jungnickel, and Frauenheim}}]{ka00}
\bibinfo{author}{\bibfnamefont{M.}~\bibnamefont{Kaukonen}},
  \bibinfo{author}{\bibfnamefont{J.}~\bibnamefont{Perajoki}},
  \bibinfo{author}{\bibfnamefont{R.~M.} \bibnamefont{Nieminen}},
  \bibinfo{author}{\bibfnamefont{G.}~\bibnamefont{Jungnickel}},
  \bibnamefont{and}
  \bibinfo{author}{\bibfnamefont{T.}~\bibnamefont{Frauenheim}},
  \bibinfo{journal}{Phys. Rev. B} \textbf{\bibinfo{volume}{61}},
  \bibinfo{pages}{980} (\bibinfo{year}{2000}).

\bibitem[{\citenamefont{Kerridge et~al.}(2004)
    \citenamefont{Kerridge, Harker, and Stoneham}
    \citenamefont{Shaw et~al.}(2005)}]{ke04}
\bibinfo{author}{\bibfnamefont{A.}~\bibnamefont{Kerridge}},
  \bibinfo{author}{\bibfnamefont{A.~H.} \bibnamefont{Harker}},
  \bibnamefont{and} \bibinfo{author}{\bibfnamefont{A.~M.}
  \bibnamefont{Stoneham}}, \bibinfo{journal}{J. Phys.: Condens. Matter}
  \textbf{\bibinfo{volume}{16}}, \bibinfo{pages}{8743} (\bibinfo{year}{2004});
\bibinfo{author}{\bibfnamefont{M.~J.} \bibnamefont{Shaw}},
  \bibinfo{author}{\bibfnamefont{P.~R.} \bibnamefont{Briddon}},
  \bibinfo{author}{\bibfnamefont{J.~P.} \bibnamefont{Goss}},
  \bibinfo{author}{\bibfnamefont{M.~J.} \bibnamefont{Rayson}},
  \bibinfo{author}{\bibfnamefont{A.}~\bibnamefont{Kerridge}},
  \bibinfo{author}{\bibfnamefont{A.~H.} \bibnamefont{Harker}},
  \bibnamefont{and} \bibinfo{author}{\bibfnamefont{A.~M.}
  \bibnamefont{Stoneham}}, \bibinfo{journal}{Phys. Rev. Lett.}
  \textbf{\bibinfo{volume}{95}}, \bibinfo{pages}{105502}
  (\bibinfo{year}{2005}).

\bibitem[{\citenamefont{Flynn and Stoneham}(1970)
          \citenamefont{Sugimoto and Fukai}(1980)
          \citenamefont{Schober and Stoneham}(1988)}]{fl70}
\bibinfo{author}{\bibfnamefont{C.~P.} \bibnamefont{Flynn}} \bibnamefont{and}
  \bibinfo{author}{\bibfnamefont{A.~M.} \bibnamefont{Stoneham}},
  \bibinfo{journal}{Phys. Rev. B} \textbf{\bibinfo{volume}{1}},
  \bibinfo{pages}{3966} (\bibinfo{year}{1970});
\bibinfo{author}{\bibfnamefont{H.}~\bibnamefont{Sugimoto}} \bibnamefont{and}
  \bibinfo{author}{\bibfnamefont{Y.}~\bibnamefont{Fukai}},
  \bibinfo{journal}{Phys. Rev. B} \textbf{\bibinfo{volume}{22}},
  \bibinfo{pages}{670} (\bibinfo{year}{1980});
\bibinfo{author}{\bibfnamefont{H.~R.} \bibnamefont{Schober}} \bibnamefont{and}
  \bibinfo{author}{\bibfnamefont{A.~M.} \bibnamefont{Stoneham}},
  \bibinfo{journal}{Phys. Rev. Lett.} \textbf{\bibinfo{volume}{60}},
  \bibinfo{pages}{2307} (\bibinfo{year}{1988}).

\bibitem[{\citenamefont{Gillan}(1987)
          \citenamefont{Voth et~al.}(1989) }]{gi87b}
\bibinfo{author}{\bibfnamefont{M.~J.} \bibnamefont{Gillan}},
  \bibinfo{journal}{J. Phys. C: Solid State Phys.}
  \textbf{\bibinfo{volume}{20}}, \bibinfo{pages}{3621}
(\bibinfo{year}{1987});
\bibinfo{author}{\bibfnamefont{G.~A.} \bibnamefont{Voth}},
  \bibinfo{author}{\bibfnamefont{D.}~\bibnamefont{Chandler}},
\bibnamefont{and}
  \bibinfo{author}{\bibfnamefont{W.~H.} \bibnamefont{Miller}},
  \bibinfo{journal}{J. Chem. Phys.} \textbf{\bibinfo{volume}{91}},
  \bibinfo{pages}{7749} (\bibinfo{year}{1989}).

\bibitem[{\citenamefont{Ram\'{\i}rez and Herrero}(1994)
      \citenamefont{Miyake et~al.}(1998) }]{ra94}
\bibinfo{author}{\bibfnamefont{R.}~\bibnamefont{Ram\'{\i}rez}}
  \bibnamefont{and} \bibinfo{author}{\bibfnamefont{C.~P.}
  \bibnamefont{Herrero}}, \bibinfo{journal}{Phys. Rev. Lett.}
  \textbf{\bibinfo{volume}{73}}, \bibinfo{pages}{126} (\bibinfo{year}{1994});
\bibinfo{author}{\bibfnamefont{T.}~\bibnamefont{Miyake}},
  \bibinfo{author}{\bibfnamefont{T.}~\bibnamefont{Ogitsu}}, \bibnamefont{and}
  \bibinfo{author}{\bibfnamefont{S.}~\bibnamefont{Tsuneyuki}},
  \bibinfo{journal}{Phys. Rev. Lett.} \textbf{\bibinfo{volume}{81}},
  \bibinfo{pages}{1873} (\bibinfo{year}{1998}).

\bibitem[{\citenamefont{Herrero et~al.}(2006)\citenamefont{Herrero,
  Ram\'{\i}rez, and Hern\'andez}}]{he06}
\bibinfo{author}{\bibfnamefont{C.~P.} \bibnamefont{Herrero}},
  \bibinfo{author}{\bibfnamefont{R.}~\bibnamefont{Ram\'{\i}rez}},
  \bibnamefont{and} \bibinfo{author}{\bibfnamefont{E.~R.}
  \bibnamefont{Hern\'andez}}, \bibinfo{journal}{Phys. Rev. B}
  \textbf{\bibinfo{volume}{73}}, \bibinfo{pages}{245211}
  (\bibinfo{year}{2006}).

\bibitem[{\citenamefont{Makarov and Topaler}(1995)}]{ma95}
\bibinfo{author}{\bibfnamefont{D.~E.} \bibnamefont{Makarov}}
\bibnamefont{and}
  \bibinfo{author}{\bibfnamefont{M.}~\bibnamefont{Topaler}},
  \bibinfo{journal}{Phys. Rev. E} \textbf{\bibinfo{volume}{52}},
  \bibinfo{pages}{178} (\bibinfo{year}{1995}).

\bibitem[{\citenamefont{Porezag et~al.}(1995)\citenamefont{Porezag, Frauenheim,
  K\"ohler, Seifert, and Kaschner}}]{po95}
\bibinfo{author}{\bibfnamefont{D.}~\bibnamefont{Porezag}},
  \bibinfo{author}{\bibfnamefont{T.}~\bibnamefont{Frauenheim}},
  \bibinfo{author}{\bibfnamefont{T.}~\bibnamefont{K\"ohler}},
  \bibinfo{author}{\bibfnamefont{G.}~\bibnamefont{Seifert}}, \bibnamefont{and}
  \bibinfo{author}{\bibfnamefont{R.}~\bibnamefont{Kaschner}},
  \bibinfo{journal}{Phys. Rev. B} \textbf{\bibinfo{volume}{51}},
  \bibinfo{pages}{12947} (\bibinfo{year}{1995}).

\bibitem[{\citenamefont{Martyna et~al.}(1996)\citenamefont{Martyna, Tuckerman,
  Tobias, and Klein}}]{ma96}
\bibinfo{author}{\bibfnamefont{G.~J.} \bibnamefont{Martyna}},
  \bibinfo{author}{\bibfnamefont{M.~E.} \bibnamefont{Tuckerman}},
  \bibinfo{author}{\bibfnamefont{D.~J.} \bibnamefont{Tobias}},
  \bibnamefont{and} \bibinfo{author}{\bibfnamefont{M.~L.} \bibnamefont{Klein}},
  \bibinfo{journal}{Mol. Phys.} \textbf{\bibinfo{volume}{87}},
  \bibinfo{pages}{1117} (\bibinfo{year}{1996}).

\bibitem[{\citenamefont{Herrero}(1997)
          \citenamefont{Noya et~al.}(1997) }]{he97}
\bibinfo{author}{\bibfnamefont{C.~P.} \bibnamefont{Herrero}},
  \bibinfo{journal}{Phys. Rev. B} \textbf{\bibinfo{volume}{55}},
  \bibinfo{pages}{9235} (\bibinfo{year}{1997});
\bibinfo{author}{\bibfnamefont{J.~C.} \bibnamefont{Noya}},
  \bibinfo{author}{\bibfnamefont{C.~P.} \bibnamefont{Herrero}},
  \bibnamefont{and}
  \bibinfo{author}{\bibfnamefont{R.}~\bibnamefont{Ram\'{\i}rez}},
  \bibinfo{journal}{Phys. Rev. Lett.} \textbf{\bibinfo{volume}{79}},
  \bibinfo{pages}{111} (\bibinfo{year}{1997}).

\bibitem[{\citenamefont{Ram\'{\i}rez}(1997)}]{ra97b}
\bibinfo{author}{\bibfnamefont{R.}~\bibnamefont{Ram\'{\i}rez}},
  \bibinfo{journal}{J. Chem. Phys.} \textbf{\bibinfo{volume}{107}},
  \bibinfo{pages}{5748} (\bibinfo{year}{1997}).

\bibitem[{\citenamefont{Odermatt et~al.}(1988)\citenamefont{Odermatt, Baumeler,
  Keller, K\"undig, Patterson, Schneider, Sellschop, Stemmet, Connell, and
  Spencer}}]{od88}
\bibinfo{author}{\bibfnamefont{W.}~\bibnamefont{Odermatt}},
  \bibinfo{author}{\bibfnamefont{H.}~\bibnamefont{Baumeler}},
  \bibinfo{author}{\bibfnamefont{H.}~\bibnamefont{Keller}},
  \bibinfo{author}{\bibfnamefont{W.}~\bibnamefont{K\"undig}},
  \bibinfo{author}{\bibfnamefont{B.~D.} \bibnamefont{Patterson}},
  \bibinfo{author}{\bibfnamefont{J.~W.} \bibnamefont{Schneider}},
  \bibinfo{author}{\bibfnamefont{J.~P.~F.} \bibnamefont{Sellschop}},
  \bibinfo{author}{\bibfnamefont{M.~C.} \bibnamefont{Stemmet}},
  \bibinfo{author}{\bibfnamefont{S.}~\bibnamefont{Connell}}, \bibnamefont{and}
  \bibinfo{author}{\bibfnamefont{D.~P.} \bibnamefont{Spencer}},
  \bibinfo{journal}{Phys. Rev. B} \textbf{\bibinfo{volume}{38}},
  \bibinfo{pages}{4388} (\bibinfo{year}{1988}).

\bibitem[{\citenamefont{Machi et~al.}(2000)
          \citenamefont{Connell et~al.}(2004) }]{ma00}
\bibinfo{author}{\bibfnamefont{I.~Z.} \bibnamefont{Machi}},
  \bibinfo{author}{\bibfnamefont{S.~H.} \bibnamefont{Connell}},
  \bibinfo{author}{\bibfnamefont{J.~P.~F.} \bibnamefont{Sellschop}},
  \bibinfo{author}{\bibfnamefont{K.}~\bibnamefont{Bharuth-Ram}},
  \bibinfo{author}{\bibfnamefont{B.~P.} \bibnamefont{Doyle}},
  \bibinfo{author}{\bibfnamefont{R.~D.} \bibnamefont{Maclear}},
  \bibinfo{author}{\bibfnamefont{J.}~\bibnamefont{Major}}, \bibnamefont{and}
  \bibinfo{author}{\bibfnamefont{R.}~\bibnamefont{Scheuermann}},
  \bibinfo{journal}{Physica B} \textbf{\bibinfo{volume}{289-290}},
  \bibinfo{pages}{468} (\bibinfo{year}{2000});
\bibinfo{author}{\bibfnamefont{S.~H.} \bibnamefont{Connell}},
  \bibinfo{author}{\bibfnamefont{I.~Z.} \bibnamefont{Machi}}, \bibnamefont{and}
  \bibinfo{author}{\bibfnamefont{K.}~\bibnamefont{Bharuth-Ram}},
  \bibinfo{journal}{Hyperf. Int.} \textbf{\bibinfo{volume}{159}},
  \bibinfo{pages}{217} (\bibinfo{year}{2004}).

\bibitem[{\citenamefont{{Van de Walle} et~al.}(1989)\citenamefont{{Van de
  Walle}, Denteneer, Bar-Yam, and Pantelides}}]{wa89}
\bibinfo{author}{\bibfnamefont{C.~G.} \bibnamefont{{Van de Walle}}},
  \bibinfo{author}{\bibfnamefont{P.~J.~H.} \bibnamefont{Denteneer}},
  \bibinfo{author}{\bibfnamefont{Y.}~\bibnamefont{Bar-Yam}}, \bibnamefont{and}
  \bibinfo{author}{\bibfnamefont{S.~T.} \bibnamefont{Pantelides}},
  \bibinfo{journal}{Phys. Rev. B} \textbf{\bibinfo{volume}{39}},
  \bibinfo{pages}{10791} (\bibinfo{year}{1989}).

\bibitem[{\citenamefont{Mehandru et~al.}(1992)\citenamefont{Mehandru, Anderson,
  and Angus}}]{me92}
\bibinfo{author}{\bibfnamefont{S.~P.} \bibnamefont{Mehandru}},
  \bibinfo{author}{\bibfnamefont{A.~B.} \bibnamefont{Anderson}},
  \bibnamefont{and} \bibinfo{author}{\bibfnamefont{J.~C.} \bibnamefont{Angus}},
  \bibinfo{journal}{J. Mater. Res.} \textbf{\bibinfo{volume}{7}},
  \bibinfo{pages}{689} (\bibinfo{year}{1992}).

\bibitem[{\citenamefont{Ramamoorthy and Pantelides}(1996)}]{ra96}
\bibinfo{author}{\bibfnamefont{M.}~\bibnamefont{Ramamoorthy}} \bibnamefont{and}
  \bibinfo{author}{\bibfnamefont{S.~T.} \bibnamefont{Pantelides}},
  \bibinfo{journal}{Phys. Rev. Lett.} \textbf{\bibinfo{volume}{76}},
  \bibinfo{pages}{267} (\bibinfo{year}{1996}).

\bibitem[{\citenamefont{Popovici et~al.}(1995)\citenamefont{Popovici, Wilson,
  Sung, Prelas, and Khasawinah}}]{po95b}
\bibinfo{author}{\bibfnamefont{G.}~\bibnamefont{Popovici}},
  \bibinfo{author}{\bibfnamefont{R.~G.} \bibnamefont{Wilson}},
  \bibinfo{author}{\bibfnamefont{T.}~\bibnamefont{Sung}},
  \bibinfo{author}{\bibfnamefont{M.~A.} \bibnamefont{Prelas}},
  \bibnamefont{and}
  \bibinfo{author}{\bibfnamefont{S.}~\bibnamefont{Khasawinah}},
  \bibinfo{journal}{J. Appl. Phys.} \textbf{\bibinfo{volume}{77}},
  \bibinfo{pages}{5103} (\bibinfo{year}{1995}).

\end{thebibliography}

\end{document}